\documentclass[journal,transmag]{IEEEtran}
\usepackage{xcolor}
\usepackage{amsmath}
\ifCLASSINFOpdf
  \usepackage[pdftex]{graphicx}
\else
  \usepackage[dvips]{graphicx}
\fi
\usepackage{siunitx}

\begin{document}
\title{Experimental Demonstration of a Spin-Wave Lens Designed with Machine Learning}

\author{\IEEEauthorblockN{Martina Kiechle\IEEEauthorrefmark{1},
Levente Maucha\IEEEauthorrefmark{2},
Valentin Ahrens\IEEEauthorrefmark{1}, 
Carsten Dubs\IEEEauthorrefmark{3}, 
Wolfgang Porod\IEEEauthorrefmark{4},
Gyorgy Csaba\IEEEauthorrefmark{2}, \\
Markus Becherer\IEEEauthorrefmark{1}, and
Adam Papp\IEEEauthorrefmark{2}}
%Wolfgang Porod\IEEEauthorrefmark{4},~\IEEEmembership{Fellow,~IEEE}}

\IEEEauthorblockA{\IEEEauthorrefmark{1}Department of Electrical and Computer Engineering, Technical University of Munich, Germany}
\IEEEauthorblockA{\IEEEauthorrefmark{2}Faculty of Information Technology and Bionics, P\'{a}zm\'{a}ny P\'{e}ter Catholic University, Budapest, Hungary}
\IEEEauthorblockA{\IEEEauthorrefmark{3}INNOVENT e.V. Technologieentwicklung, Jena, Germany}
\IEEEauthorblockA{\IEEEauthorrefmark{4}Department of Electrical Engineering, Notre Dame, University of Notre Dame, IN, USA}% <-this % stops an unwanted space

\thanks{Manuscript received June 30, 2022; revised .., 2022. 
Corresponding author: Adam Papp (email: papp.adam@itk.ppke.hu)}}

\markboth{IEEE Magnetics Letters, Vol. xx, 2022}%
{Shell \MakeLowercase{\textit{et al.}}: Experimental Demonstration of a Spin-wave Lens Designed with Machine Learning}

\IEEEtitleabstractindextext{%
\begin{abstract}
We present the design and experimental realization of a device that acts like a spin-wave lens i.e., it focuses spin waves to a specified location. The structure of the lens does not resemble any conventional lens design, it is a nonintuitive pattern produced by a machine learning algorithm. As a spin-wave design tool, we used our custom micromagnetic solver 'SpinTorch' that has built-in automatic gradient calculation and can perform backpropagation through time for spin-wave propagation. The training itself is performed with the saturation magnetization of a YIG film as a variable parameter, with the goal to guide spin waves to a predefined location. We verified the operation of the device in the widely used mumax3 micromagnetic solver, and by experimental realization. For the experimental implementation, we developed a technique to create effective saturation-magnetization landscapes in YIG by direct focused-ion-beam irradiation. This allows us to rapidly transfer the nanoscale design patterns to the YIG medium, without patterning the material by etching. We measured the effective saturation magnetization $M_\mathrm{eff}$ corresponding to the FIB dose levels in advance and used this mapping to translate the $M_\mathrm{eff}$ values of the design to the required dose levels. 
Our demonstration serves as a proof of concept for a workflow that can be used to realize more sophisticated spin-wave devices with complex functionality, e.g., spin-wave signal processors, or neuromorphic devices.
\end{abstract}

\begin{IEEEkeywords}
spin waves, spin-wave optics, inverse-design magnonics, machine learning.
\end{IEEEkeywords}}

\maketitle

\IEEEdisplaynontitleabstractindextext

\IEEEpeerreviewmaketitle

\section{Introduction}
\label{sec:introduction}
\IEEEPARstart{T}{here} is an undisputed need to improve the efficiency (especially the power efficiency) of computing devices. The exponential growth of available computing power -- from personal devices to large data centers -- is limited by the energy consumption of the computing architectures. There is also a consensus that the power efficiency of digital, Boolean devices cannot be improved by simply replacing CMOS transistors with some different switching element. 

Magnons can carry and process information at very high (several \SI{10}{\giga\hertz}) bandwidths, dissipate very little energy (attojoules), and hence they are a potential candidate for beyond-Moore devices. Information processing does not necessarily have to be \emph{digital}, instead wave-based,  neuromorphic, and non-Boolean building blocks are widely seen as the most promising  unconventional computing paradigms with CMOS being the most advanced digital technology platform without seriously competitive successors. 

Spin-wave optical elements replicate the functions of optical computing blocks and provide a pathway to signal processing and computing functions. This motivates our work to design compact spin-wave optics elements, which are beyond the realm of classical optics. Magnonics went through an intense development in the past decade -- in particular, it became possible to experimentally demonstrate devices that are magnonic versions of known optical structures in the spin-wave domain, as proposed by \cite{csaba2014spin}. The majority of magnonic devices demonstrated so far aims to replicate the behavior of the elements of classical optics. Nano-optical devices \cite{molesky2018inverse}, however, may turn out to be a better fit for the capabilities and limitations of magnonics. One advantage nanophotonic devices have over classical optics is that they typically yield higher functionality in a smaller footprint -- this could be a crucial benefit in magnonics, where damping limits device size. This observation motivates our work to design magnonic devices that are not derived from the elements of classical optics.

Nanophotonic devices are most often complex, non-intuitive scattering structures, often designed by machine learning methods \cite{genty2021machine}. We follow this route, by using our recently developed SpinTorch code \cite{SpinTorch} to design a linear scatterer that acts like a focusing lens, despite not at all resembling one in light of its structure. 

Recently, multiple approaches emerged in the magnonics community for the inverse design of spin-wave scattering devices \cite{wang2021inverse,papp2021nanoscale}. These designs show a possible path towards neuromorphic computing with spin waves, magnonic signal processing and complex logical gates. However, the feasibility of experimental realization of the presented devices is not satisfactorily addressed in these works. Here we use a simple device -- a lens -- for the demonstration of a workflow that includes both the inverse design of the device, and a fabrication method to implement it.
\section{Design of a Spin-Wave Lens with Machine Learning}
\label{sec:design}
In contrast to using fundamental design principles for spin-wave devices (e.g. optical design methods), the use of machine learning allows us to discover non-intuitive design geometries and may result in improved performance, more compact footprint, and higher level of integration compared to optical systems built from discrete components. In this work, we use a lens (focusing to a specified output location) as an example of such a design to demonstrate the operation of our micromagnetic inverse-design tool, and also the feasibility of experimental realizations of such designs by our experimental technique. 
\subsection{SpinTorch: a Micromagnetic Solver with Machine Learning Integrated}
We have developed and implemented a micromagnetic simulator, SpinTorch \cite{SpinTorch}, with built-in gradient-based optimization capability. We have previously demonstrated that this algorithm allows us to design spin-wave scatterers that can perform various computing tasks \cite{papp2021nanoscale}. More specifically, we have shown that these scatterer designs can perform similar tasks as neural networks, e.g. solve classification problems, linear transformations, and even nonlinear functions if spin waves are used in the nonlinear amplitude regime. 

SpinTorch is built on Pytorch \cite{PyTorch}, a machine-learning framework that is widely used to build and train neural networks. It has automatic backpropagation capability, which is realized by building a computational graph of the neural network during the forward run. In our case, this means backpropagation through time for spin-wave propagation. In SpinTorch, the trainable parameters can be any parameters of the micromagnetic simulation, e.g. spatial distribution of the external field, or saturation magnetization -- the latter is used in this work. The desired functionality is defined through inputs and outputs, which can be transducers on the propagation medium. Training -- or inverse design -- is achieved in a series of epochs, during which spin waves are launched, interference patterns and output signals are calculated, and based on the gradients the design is updated until satisfactory performance is achieved.
\subsection{Lens design}
A relatively simple but nevertheless powerful application of SpinTorch is the inverse design of spin-wave optical elements. For demonstration of our methods, we chose to design a lens, i.e. a device that focuses a wavefront to a specified location. A classical lens in optics achieves focusing by refracting waves on the lens surfaces: its operation is determined by the lens curvature and the relative refractive index of its material. In our machine-learning approach, the refractive index is varied pointwise in the specified design region by the algorithm. This 'distributed-parameter' approach enables better utilization of space, and control of aberrations in the system by the definition of an appropriate objective function.

\begin{figure}[ht]
    \centering
    \includegraphics[width=0.48\textwidth]{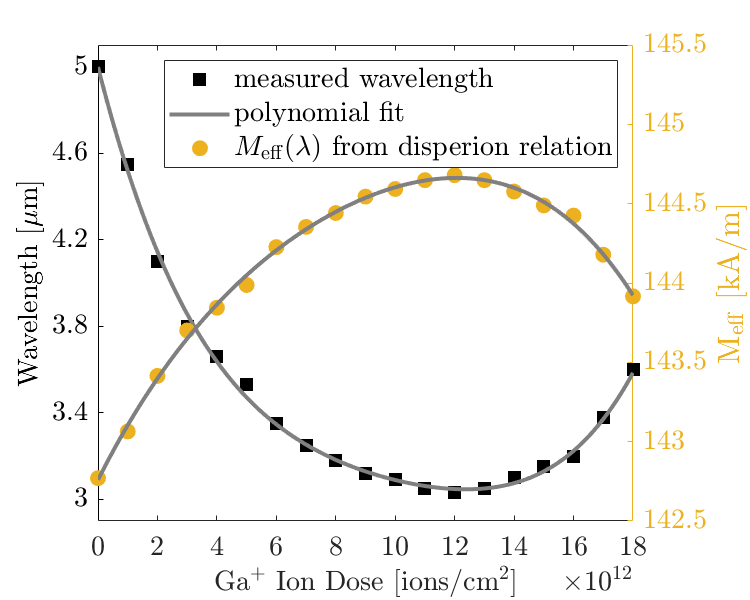}
    \caption{Nonlinear trend of the on-dose-dependent change in YIG. The modified wavelengths $\lambda_\mathrm{FIB}$ were measured in 38x38\,µm regions irradiated with the respective ion dose. Subsequently, the effective magnetization change was calculated from the respective $\lambda_\mathrm{FIB}$ and the spin-wave dispersion relation. The largest $\Delta M_\mathrm{eff}$~=~144.7\,kA/m is used as the basis for the training.}
    \label{fig:dose_lambda}
\end{figure}

In order to modify the refractive index of YIG for spin waves, we used FIB irradiation experimentally. In simulations and training, we modelled the effect of irradiation by a change in effective saturation magnetization ($M_\mathrm{eff}$). The exact mapping between irradiation dose and $M_\mathrm{eff}$ was extracted experimentally through measurement of the wavelength (Fig. \ref{fig:dose_lambda}). We defined a 50\,µm by 50\,µm design region in the YIG film where the algorithm was allowed to modify the intrinsic effective magnetization between the intrinsic value $M_\mathrm{eff,0}$~=~142.8\,kA/m, and the maximum achievable by FIB irradiation $M_\mathrm{eff,max}$~=~144.7\,kA/m. 

\begin{figure*}[ht]
    \centering
    \includegraphics[width=\textwidth]{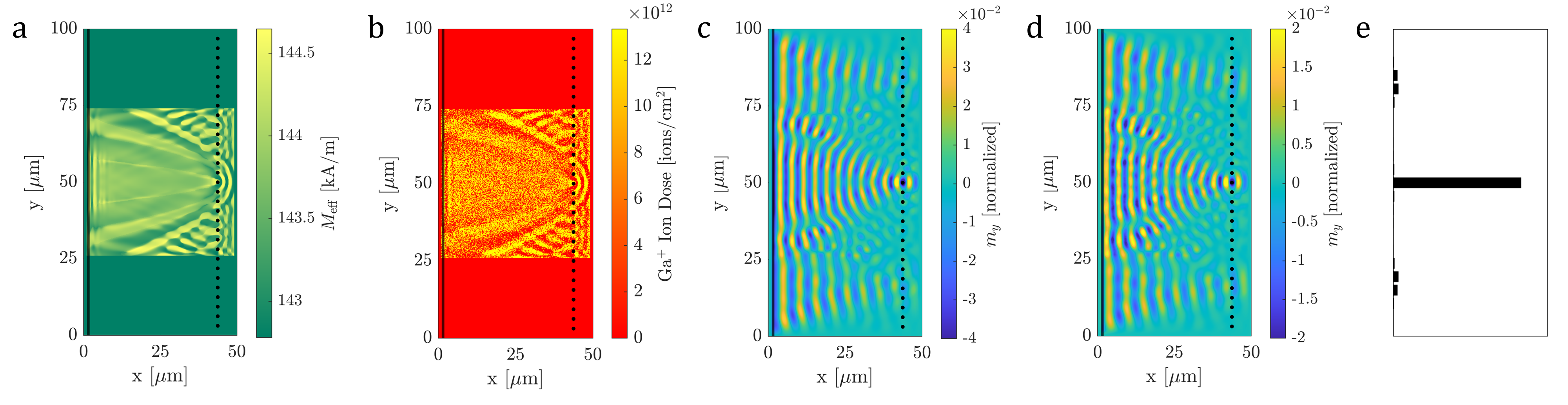}
    \caption{Sequence of the lens design flow. (a) Machine-learned saturation-magnetization pattern trained to focus spin waves to the center output. (b) Binarized scatterer pattern with increased resolution to match the FIB spot size and mapped to the corresponding ion dose values. (c) Spin-wave interference pattern calculated by SpinTorch after finished training (corresponding to the pattern in (a)). (d) Mumax3 simulated spin-wave propagation in the binary pattern from (b). (e) Spin-wave intensities on the defined outputs in c). }
    \label{fig:simulations}
\end{figure*}

\begin{figure*}[ht]
    \centering
    \includegraphics[width=\textwidth]{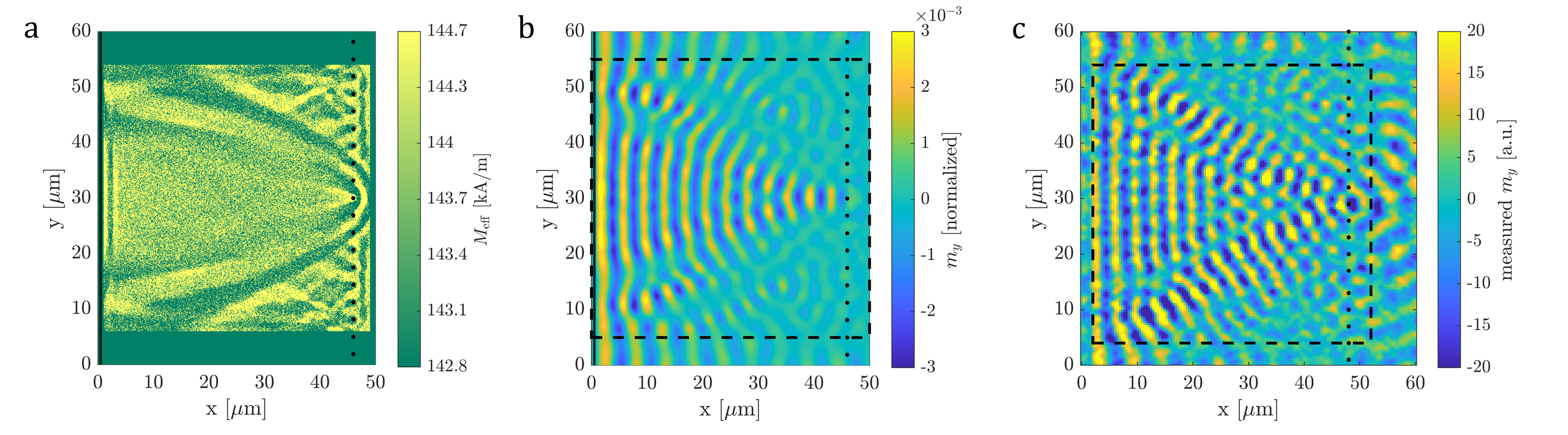}
    \caption{Focusing of spin waves by a machine-learning-designed magnetization pattern. (a) SpinTorch uses the two values $M_\mathrm{eff,0}$ and $M_\mathrm{eff,FIB}$ as training parameters, and constructs a binary non-trivial saturation magnetization map by the inverse-design algorithm. (b) plane wave propagation through the pattern in (a) simulated in mumax3. (c) Measured spin wave waveform in the FIB irradiated magnetization pattern, showing the wavefront focusing to the center output. The dashed rectangle corresponds to the 50x50\,µm design area used for the training.}
    \label{fig:center}
\end{figure*}

\begin{figure*}[ht]
    \centering
    \includegraphics[width=\textwidth]{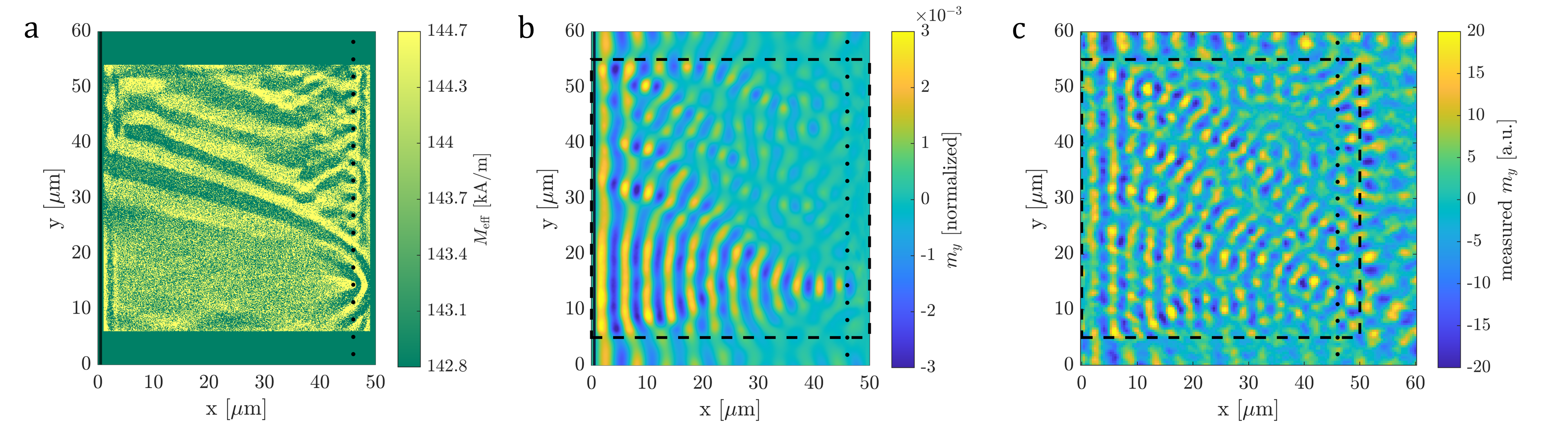}
    \caption{Offside focusing of spin waves to different output. (a) The binary magnetization pattern created in spinTorch. (b) mumax3 simulation of a plane wave ($\lambda_0$~=~5\,µm) excited on the left, traveling through the binary magnetization pattern shown in (a). The wavelength changes due to the locally changing magnetization, which shapes the focusing to the side as originally intended. (c) trMOKE image of the FIB irradiated pattern with the wavefront focusing to the lower output.}
    \label{fig:offside}
\end{figure*}

We defined an artificial line source adjacent to the design region, which excites a linear spin-wave wavefront. Outputs were placed on the opposite side, circular regions with diameters of 1\,µm arranged in the focal plane with 3,125\,µm separation in between. Outside the design region, we included an absorbing boundary layer, which absorbs spin waves by a smoothly increasing damping coefficient. On the lateral sides of the design area, we also included 25\,µm padding on both sides (truncated in Fig.~\ref{fig:center} and Fig.~\ref{fig:offside}, but shown in Fig. \ref{fig:simulations}), where normal YIG parameters were assumed. This is required because in the experiment the excited wavefront is much longer than the irradiated region, so diffraction of waves from the neighboring regions should be considered for an optimal design.

For the simulation we used a damping coefficient value $\alpha_\mathrm{YIG}\,=\,7.9\times10^{-4}$, exchange coefficient $A_\mathrm{exch}~=~3.65$\,pJ/m, gyromagnetic ratio $\gamma_\mathrm{LL}~=~1.7595\times10^{11}$\,rad/Ts. A YIG-film thickness of 69\,nm and 2D discretization of 100\,nm by 100\,nm were used. The excitation frequency was set to 3\,GHz, the bias field to 282.5\,mT (out-of-plane direction), which resulted in a spin-wave wavelength of $\lambda_0$~=~5\,µm in the unirradiated regions, and $\lambda_\mathrm{FIB}$~=~3\,µm where maximum FIB dose was applied.

We run the optimization in SpinTorch for 14 epochs, where each epoch consists of a forward micromagnetic simulation, a backward gradient calculation, and updating the design parameters, i.e. the saturation magnetization distribution (the result of the training is shown in Fig. \ref{fig:simulations}a). Due to technical limitations of our FIB instrument, the final saturation magnetization pattern -- which contained continuous values -- had to be converted to binary values before irradiation (Fig. \ref{fig:simulations}b). For this, we increased the discretization resolution by a factor of two in both lateral dimensions, and assigned a binary value to every pixel with a probability that was linearly mapped from the designed pattern (this process can be referred to as stochastic halftoning). Here we assumed that since the pixel size is almost two orders of magnitude smaller than the spin-wave wavelength, spin waves will 'see' only the average values.

\subsection{Design Verification with mumax3}
As a validation step of the training, we repeated the micromagnetic simulations in mumax3 \cite{mumax} with the final design pattern. Due to limitations of mumax3 (limited number of distinct $M_s$ values) we used the binarized pattern in these simulations, this way also verifying our assumptions with the binary mapping. Fig.~\ref{fig:simulations}b and d show the binary pattern and the resulting interference pattern, respectively. A comparison with Fig.~\ref{fig:simulations}c shows good agreement between the two simulations, the slight noise in the mumax3 simulation can be attributed to the non-exact stochastic approach taken in the binarization of the pattern.
\section{Experimental Realization}
\label{sec:experiments}
Direct focused-ion-beam (FIB) irradiation of yttrium-iron-garnet (YIG) films has been recently demonstrated as an efficient fabrication method for magnetization landscapes with nanoscale precision \cite{kiechle2022spin}. For the experimental arrangement, a 69 nm thin YIG film with coplanar microwave antennas is used to excite spin waves with out-of-plane magnetic-field bias. The YIG film was fabricated with liquid phase epitaxy \cite{Dubs_YIG}, and its magnetic parameters obtained from ferromagnetic-resonance (FMR) measurements are $M_\mathrm{eff,0}$~=~142.8~kA/m and $\alpha$~=~0.0004. The 2D spin-wave patterns were recorded with longitudinal time-resolved magneto-optical Kerr effect (trMOKE) microscopy. 
A microwave frequency of $f$~=~3\,GHz at a power of  $P$~=~-10\,dBm and a bias field of approximately $H_\mathrm{dc}$~=~282\,mT resulted in a spin-wave wavelength of $\lambda_0$~=~5\,µm, which was used as a basis for the simulations and experiments. 
\subsection{Engineering the Effective Magnetization in YIG via FIB}
The change of the effective magnetization is ion-dose dependent, and the degree of change is measured by the wavelength change ($\lambda_0$ vs. $\lambda_\mathrm{FIB}$) of spin waves traveling through irradiated regions (38x38\,µm in size) of various ion doses. In Fig.~\ref{fig:dose_lambda} the ion dose vs. wavelength profile is shown (numerically on the left axis). It becomes clear that the trend is non-linear, and the trend turns around after achieving a minimal wavelength (doses beyond this value have little practical importance, since the damping strongly increases above this dose). We note that the profile shape changes when a different $\lambda_0$ is used, which is due to the non-linear dispersion relation. Additionally, the $M_\mathrm{eff}$ change with respect to the ion dose is presented, which was calculated using the dispersion relation.   
The ratio of $\lambda_0$/$\lambda(M_\mathrm{eff,FIB})$ can be modeled as the adjustable refractive index $n$ in FIB-irradiated regions, given that $n$~=~1 in intrinsic YIG. For the training of the binary lens patterns it is instructive to use the maximum achievable magnetization difference, hence the intrinsic value $M_\mathrm{eff,0}$~=~142.8 kA/m and largest $M_\mathrm{eff,max}$~=~144.7 kA/m were used. The magnetization distribution obtained from the training in Fig.~\ref{fig:center}a and Fig.~\ref{fig:offside}a can directly be used as an irradiation image in the FIB tool, and the yellow regions were irradiated at an ion dose of $1.3~\cdot~10^{13}$~ions/cm$^2$ with a pixel size of 50 nm (equivalent to the cell size used in the mumax3 simulations).
\subsection{Experimental Verification of the Lens Design}
For the imaging of spin wave propagation in YIG, we use an in-house developed time-resolved optical Kerr microscope (trMOKE). This -- alongside with other optical -- imaging technique can directly monitor spin waves in k-space and hence, provides the spatio-temporal environment needed for comparison with the simulations. The FIB irradiated patterns are optically invisible and no material in the YIG film is removed, merely the ion implantation causes a structural change of the crystal that changes the effective magnetization on the nanoscale, thus propagating spin waves with micrometer wavelengths face negligible discontinuities. For the purpose of demonstrating the design flow, i.e. the training of the $M_\mathrm{eff}$ map, its simulation with plane waves in mumax3 and eventually the experimental verification of the FIB irradiated magnetization map, we show two lens-like operating elements that aim to focus spin waves to different locations. In the case of focusing to the center output, the trMOKE image (Fig.~\ref{fig:center}c) of propagating spin waves through the FIB irradiated magnetization pattern resembles the simulated scene, with the associated wave traces clearly visible. As for the offside focusing shown in Fig.~\ref{fig:offside}c, the individual wave paths can still be recognized and the focus is in the right location, although not at highest intensity and the image itself has a much noisier background. There are many potential reasons for such imperfections, such as fluctuations in the ion beam current, resulting in an ion dose deviation, and therefore producing a different magnetization value. Using precise parameters is crucial, a small deviation in local magnetization changes can have a large impact on spin-wave interactions while propagating through. Aside from this, more common issues in spin-wave experiments such as imperfections in the YIG film or unwanted interference with waves entering the pattern from the side can distort wave form in the areas of interest. Magnetic damping is considerably low in YIG thin films, and spin-wave propagation in the few tens of micrometer areas shown in this work can propagate without much amplitude decrease. However, FIB irradiation adds a somewhat moderate damping contribution that will increase with the ion dose, which we did not take into account for the training. 

Nevertheless, we observe good agreement between the simulated and experimentally fabricated magnetization patterns, and they show the intended behavior, i.e., focusing spin waves to a predefined location. This proves that SpinTorch can successfully design transfer functions/functional areas for spin waves and FIB can be used as an effective prototyping and fabrication tool for any magnetization pattern within the available range at nanoscale precision. Combining those two methods provides a powerful toolbox to design non-trivial spin-wave computing elements.   
\section{Conclusion}
We demonstrated a general design method for spin-wave optics through the specific example of a lens. We experimentally verified our designs via direct-FIB irradiation of YIG films. Our results show a good agreement between simulation and experiment, we hope that this demonstration could ingnite interest in inverse magnonics as a possible route towards nanoscale signal processors and computing devices. 

While there have been numerous demonstrations of magnonic logic \cite{wang2020magnonic,vogt2014realization,sadovnikov2016frequency} and optically-inspired spin-wave computing \cite{dai2020focusing}, none of these designs could overcome the challenges of the field, first of all the inefficiency of the transducers combined with extremely small signal levels and moderately high damping. We believe that our results could be used to tackle some of these issues by increasing the complexity of the functions that can be realized in the spin-wave domain before readout or amplification of the signal becomes necessary.

\section*{Acknowledgment}

The authors want to thank all staff members and researchers working in the lab facilities of ZEIT\textsuperscript{lab}, an organizational unit of the Department of Electrical and Computer Engineering at TUM. We highly appreciate the funding from the German Research Foundation (DFG No.~429656450 and DFG No.~271741898) and the German Academic Exchange Service (DAAD, No.~57562081). 
Adam Papp received funding from the PPD research program of the Hungarian Academy of Sciences.

\bibliographystyle{./IEEEtran}
\bibliography{./IEEEabrv,./References}

%\newpage
%\begin{IEEEbiography}{Martina Kiechle}
%\end{IEEEbiography}
%\begin{IEEEbiography}{Gyorgy Csaba}
%\end{IEEEbiography}
%\begin{IEEEbiography}{Markus Becherer}
%\end{IEEEbiography}
%\begin{IEEEbiography}{Adam Papp}
%\end{IEEEbiography}

%\begin{IEEEbiographynophoto}{Gyorgy Csaba}

\end{document}